\documentclass[journal]{IEEEtran}
\usepackage{graphicx,amssymb,amsmath,amsthm}
\usepackage{multicol}
\usepackage[noadjust]{cite}
\usepackage{setspace}
\usepackage{stfloats}
\usepackage{midfloat}
\usepackage[normal]{threeparttable}
\usepackage{flushend,cuted}
\usepackage{cuted}
\usepackage{cases}
\usepackage{bm}
\usepackage{textcomp}
\usepackage{latexsym,bm}
\usepackage{booktabs,changebar}
\usepackage{xcolor}
\usepackage[top=0.5in, bottom=0.5in, left=0.5in, right=0.5in]{geometry}

\newtheorem{them}{Theorem}
\newtheorem{coro}{Corollary}

\newtheorem*{remark}{Remark}

\IEEEoverridecommandlockouts
\begin{document}
\title{On the Multivariate Gamma-Gamma ($\Gamma \Gamma$) Distribution with Arbitrary Correlation and Applications in Wireless Communications
}

\author{Jiayi~Zhang,~\IEEEmembership{Member,~IEEE,}
        Michail~Matthaiou,~\IEEEmembership{Senior~Member,~IEEE,}
        George~K.~Karagiannidis,~\IEEEmembership{Fellow,~IEEE,}
        and~Linglong Dai,~\IEEEmembership{Senior~Member,~IEEE}

\thanks{
Copyright (c) 2015 IEEE. Personal use of this material is permitted. However, permission to use this material for any other purposes must be obtained from the IEEE by sending a request to pubs-permissions@ieee.org.
}
\thanks{
This work was supported by the National Key Basic Research Program of China (No. 2013CB329203), National Natural Science Foundation of China (Grant No. 61201185), National High Technology Research and Development Program of China (Grant No. 2014AA01A704), and China Postdoctoral Science Foundation (No. 2014M560081).}%

\thanks{J. Zhang and L. Dai are with the Department of Electronic Engineering as well as Tsinghua National Laboratory of Information Science and Technology (TNList), Tsinghua University, Beijing 100084, P. R. China (e-mail: \{jiayizhang, daill\}@tsinghua.edu.cn).}
\thanks{M. Matthaiou is with the School of Electronics, Electrical Engineering and Computer Science, Queen's University Belfast, Belfast, BT3 9DT, U.K. and with the Department of Signals and Systems, Chalmers University of Technology, Gothenburg SE-412 96, Sweden (e-mail: m.matthaiou@qub.ac.uk).}
\thanks{G. K. Karagiannidis is with the Department of Electrical and Computer Engineering, Khalifa University, PO Box 127788, Abu Dhabi, UAE and with the Department of Electrical and Computer Engineering, Aristotle University of Thessaloniki, 54 124, Thessaloniki, Greece (e-mail: geokarag@ieee.org).}

}

\maketitle
\vspace{-2cm}
\begin{abstract}
The statistical properties of the multivariate Gamma-Gamma ($\Gamma \Gamma$) distribution with arbitrary correlation have remained unknown. In this paper, we provide analytical expressions for the joint probability density function (PDF), cumulative distribution function (CDF) and moment generation function of the multivariate $\Gamma \Gamma$ distribution with arbitrary correlation. Furthermore, we present novel approximating expressions for the PDF and CDF of the sum of $\Gamma \Gamma$ random variables with arbitrary correlation. Based on this statistical analysis, we investigate the performance of radio frequency and optical wireless communication systems. It is noteworthy that the presented expressions include several previous results in the literature as special cases.
\end{abstract}


\IEEEpeerreviewmaketitle
\section{Introduction}
In practical wireless communications, the electromagnetic signal experiences composite small-scale fading and shadowing simultaneously. Over the past decades, several works have focused on the performance analysis of wireless systems over composite fading channels \cite{simon2005digital,Zhang2012performance,paris2014statistical}.
On a different note, diversity techniques are widely used to reduce the effects of multipath fading. If the antennas are sufficiently separated, it is reasonable to assume that signals received by different antennas are independent. However, this assumption is intimately crude for systems with closely spaced antennas, such as mobile phones. In general, correlation between channels results in a degradation of the diversity gain. Therefore, it is important to quantify rigorously the effects of correlation in real-life practical scenarios.

Recently, the Gamma-Gamma ($\Gamma \Gamma$) distribution has been proposed as a general and mathematically tractable composite fading model \cite{peppas2011multivariate,bithas2009bivariate,chatzidiamantis2011distribution}. The $\Gamma \Gamma$ distribution is equivalent to the squared generalized-$K$ distribution and includes the squared $K$ distribution and double-Rayleigh distribution as special cases. In radio frequency (RF) communications, the performance of diversity systems over $\Gamma \Gamma$ fading channels has been analyzed in \cite{peppas2011multivariate,chatzidiamantis2011distribution}. In particular, \cite{peppas2011multivariate} investigated the outage probability (OP) of selection combining (SC) receivers, as well as the average bit error rate (BER) of maximal ratio combining (MRC) receivers operating over exponentially correlated distributed fading channels. Furthermore, a simple approximation for the distribution of the sum of independent distributed $\Gamma \Gamma$ random variables (RVs) was provided in \cite{chatzidiamantis2011distribution}. However, all these works are based on the assumption of independent or exponentially correlated $\Gamma \Gamma$ RVs. Furthermore, the $\Gamma \Gamma$ distribution has been used in the performance investigation of free-space optical (FSO) links over atmospheric turbulence conditions \cite{uysal2004error,chatzidiamantis2011distribution,zhang2015unified,zhang2015ergodic}. These previous results are not theoretically attractive when the underlying FSO links are not independently distributed. To the best of the authors' knowledge, a statistical analysis of \emph{the multivariate $\Gamma \Gamma$ distribution with arbitrary correlation} is still not available in the literature.

Motivated by the above discussion, this paper makes the following specific contributions:
\vspace{-2mm}
\begin{itemize}
\item Using the Green's matrix approximation \cite{karagiannidis2003efficient}, we derived new analytical expressions for the probability density function (PDF), cumulative distribution function (CDF) and moment generation function (MGF) of the joint $\Gamma \Gamma$ distribution with arbitrary correlation. We point out that the presented results encompass several previously known results as special cases (e.g., those in \cite{bithas2009bivariate,bithas2006performance,peppas2011multivariate}).

\item Efficient approximations to the PDF and CDF of the sum of arbitrarily correlated $\Gamma \Gamma$ RVs are derived for integer values of fading severity $m$. The resulting expressions can be easily and quickly evaluated by using a recursive formula.

\item Furthermore, in order to reveal the importance of the proposed statistical formulations, we study the performance of SC and MRC receivers over arbitrarily correlated generalized-$K$ fading channels.

\item Finally, in the context of FSO communication systems with spatial diversity, we analyze the BER performance when strong turbulence channels is assumed. The derived expression extends the results on independent and exponentially correlated $\Gamma \Gamma$ fading channels in \cite{peppas2011multivariate}.
\end{itemize}

\emph{Notations}: We use upper and lower case boldface to denote matrices and vectors, respectively. The expectation is given by ${\text E}\left(\cdot\right)$. The norm of a vector is given by $||\cdot||$, while the matrix determinant by $|\cdot|$. The Hermitian operation is defined as $\left(\cdot\right)^\dag$. The set of integer numbers is expressed via $\mathbb{Z}$. Finally, let $ {\pmb{X}}_{i,j}$ denote the $\left(i,j\right)$-th element of the matrix $\pmb{X}$.

\section{Multivariate distribution with arbitrary correlation}\label{se:2}
\subsection{Multivariate Nakagami-$m$ Distribution with Arbitrary Correlation}
Let $r$ be a Nakagami-$m$ RV, whose PDF is given by \cite{karagiannidis2003efficient}
\begin{align}\label{eq:Naka_pdf}
f_r\left( r \right) = \frac{{2{r^{2m - 1}}}}{{\Gamma \left( m \right){\Omega ^m}}}\exp \left( { - \frac{{{r^2}}}{\Omega }} \right), \;\;\; r \geq 0,
\end{align}
where $\Gamma(\cdot)$ represents the Gamma function \cite[Eq. (8.310.1)]{gradshtein2000table}, $\Omega \triangleq {{\text E} (r^2) }/m$ with ${\text E}(r^2)$ being the average power, and $m \geq 1/2$ denotes the fading severity. Note that \eqref{eq:Naka_pdf} is another representation of the classical formula for the single Nakagami-$m$ PDF. Without loss of generality, let $u$ be a Gamma RV which represents shadowing and it is assumed in the following that ${\text E} (u^2) = 1$. Then, the PDF of $u$ can be expressed as \cite{uysal2004error}
\begin{align}\label{eq:gamma_pdf}
f_u\left( u  \right) = \frac{{{\beta ^\beta }}}{{\Gamma \left( \beta  \right)}}{u ^{\beta  - 1}}{e^{ - \beta u }}, \;\;\; u,   \beta \geq 0,
\end{align}
where $\beta$ is a channel parameter related to the effective number of discrete scatterers.
Let ${\mathbf{y}}_1, {\mathbf{y}}_2, \ldots, {\mathbf{y}}_{2m}$ be $N$-dimensional column vectors, which are independent of each other and normally distributed with zero means and arbitrary correlation matrix $\pmb{\Sigma}$ which is defined as ${\pmb{\Sigma}}_{i,j}=1$ for $i=j$, and ${\pmb{\Sigma}}_{i,j}=\rho_{i,j}$ for $i\neq j$, where $0\leq \rho_{i,j} <1$ is the correlation coefficient and $1 \leq {i,j} \leq N$. Here we assume that all vectors have the same correlation matrix $\pmb{\Sigma}$.

Now, let ${\mathbf{x}}_1, {\mathbf{x}}_2, \ldots, {\mathbf{x}}_N$ be $2m$-dimensional column vectors, with $\mathbf{x}_n$ composed of the $n$-th components of the ${\mathbf{y}}_i$ and the norm of ${\mathbf{x}}_n$ being $r_n$, i.e., $r_n \triangleq ||{\mathbf{x}}_n||$. Then, $r_i$ is a Nakagami-$m$ RV. We recall that the joint PDF of $N$ Nakagami-$m$ RVs, $\pmb{r} \triangleq [r_1, r_2, \ldots, r_N]$, with arbitrary correlation and identically distributed is given by \cite[Eq. (2)]{karagiannidis2003efficient} \cite[Eq. (9)]{alexandropoulos2009new}, as follows
\begin{align}\label{eq:arbitrary_Nakagami_PDF}
f_{\pmb{r}}\left( \pmb{r} \right)&= \frac{{{{\left| \pmb{W} \right|}^m}{2^N}r_1^{m - 1}r_N^m}}{{{\Omega ^{N+m - 1}}\Gamma \left( m \right)}}{e^{ - \frac{{{p_{N,N}}r_N^2}}{\Omega }}} \prod \limits_{n = 1}^{N - 1} \Bigg[ {{\left| {{p_{n,n + 1}}} \right|}^{ 1-m}}{r_n}\notag \\
&\times{e^{ - \frac{{{p_{n,n}}r_n^2}}{\Omega }}}{I_{m - 1}}\left( {\frac{{2\left| {{p_{n,n + 1}}} \right|}}{\Omega }{r_n}{r_{n + 1}}} \right) \Bigg],
\end{align}
where $\pmb{W}$ denotes the inverse of $\pmb{\Sigma}$ with elements $p_{i,j}$, and $I_v(\cdot)$ is the modified Bessel function of the first kind with order $v$ \cite[Eq. (8.406)]{gradshtein2000table}.
\begin{remark}
{Note that, although the parameter $2m$ seems to be restricted as a positive integer, \eqref{eq:arbitrary_Nakagami_PDF} can be used for any positive values of $m$ not less than $1/2$. This is because it satisfies the necessary and sufficient conditions to be a joint distribution function, as it was emphasized in \cite{karagiannidis2003efficient}.}
\end{remark}

The joint PDF expression (\ref{eq:arbitrary_Nakagami_PDF}) requires the matrix $\pmb{W}$ to have the tridiagonal property. However, in the general case, the inverse of $\pmb{\Sigma}$ is not a tridiagonal matrix. Therefore, we need to approximate $\pmb{\Sigma}$ with a Green's matrix $\pmb{C}$ \cite{nabben2001green}, which has the form
\begin{align}\label{eq:green_matrix}
\pmb{C} = \left[ {\begin{array}{*{20}{c}}
{{\zeta_1}{\vartheta_1}}&{{\zeta_1}{\vartheta_2}}&{\ldots}&{{\zeta_1}{\vartheta_N}}\\
{{\zeta_1}{\vartheta_2}}&{{\zeta_2}{\vartheta_2}}&{\ldots}&{{\zeta_2}{\vartheta_N}}\\
{\vdots}&{\vdots}&{\ddots}&{\ldots}\\
{{\zeta_1}{\vartheta_N}}&{{\zeta_2}{\vartheta_N}}&{\ldots}&{{\zeta_N}{\vartheta_N}}
\end{array}} \right],
\end{align}
where $\zeta_i$ and $\vartheta_i$, are two sequences of real numbers. In order to determine $\pmb{C}$, nonlinear methods (e.g., Levenberg Marquardt, quasi Newton, and conjugate
gradient) may be used to solve the equation $\pmb{C}=\pmb{\Sigma}$, i.e., $\zeta_i \vartheta_i={{\pmb{\Sigma}}_{i,i}}=1$ and $\zeta_i \vartheta_j={\pmb{\Sigma}}_{i,j} = \rho_{i,j}$.  {The accuracy of the Green's matrix approximation technique will be verified in Section \ref{se:5}. In any case, its computational complexity, determined by the dimension of the correlation matrix, is given by $\mathcal{O}(N^2)$.}

\subsection{Multivariate Gamma Distribution with Arbitrary Correlation}
It is well known that a Nakagami-$m$ RV is the square root of a Gamma RV. Furthermore, it has been numerically validated in \cite{zhang2004simulation} that the correlation coefficients of Gamma RVs can be taken identical to the corresponding coefficients of Nakagami-$m$ RVs with satisfactory accuracy.
Performing $N$ RV transformations onto (\ref{eq:arbitrary_Nakagami_PDF}), the joint PDF of multiple Gamma RVs, $\pmb{\omega} \triangleq [\omega_1, \omega_2, \ldots, \omega_N]$ with {$\omega_i=r_i^2=||{\mathbf{x}}_i||^2$} for $i=1,\ldots,N$, can be expressed as
\begin{align}
&f_{\pmb{\omega}}\left( {\pmb{\omega}}\right)=  \frac{{{{\left| \pmb{W} \right|}^m}}{e^{ - \sum\limits_{n = 1}^N {\frac{{{p_{n,n}}{\omega _n}}}{\Omega }} }}}{{ \Gamma \left( m \right)}}\sum\limits_{{i_1}, \cdots ,{i_{N - 1}} = 0}^\infty \omega _1^{m + {i_1} - 1}\omega _N^{m + {i_{N-1}} - 1}  \notag \\
&\times{\Omega ^{ - \!Nm \!- \!2\sum\limits_{j = 1}^{N \!- \!1} {{i_j}} }}  \prod\limits_{n = 1}^{N \!-\!\! 1} \prod\limits_{j = 2}^{N \!-\! 1} \omega _j^{m \!+\! {i_{j \!-\! 1}} \!+\! {i_j} \!-\! 1} \prod\limits_{n = 1}^{N \!-\! 1} {\left[ {\frac{{{\left| {{p_{n,n \!+\! 1}}} \right|}^{   2 {i_n}  }}}{{{i_n}!\Gamma \left( {m + {i_n}} \right)}}} \right]}  . \label{eq:arbitrary_gamma_PDF1}
\end{align}
To derive (\ref{eq:arbitrary_gamma_PDF1}), we have used the infinite series representations of $I_v(\cdot)$ \cite[Eq. (8.445.1)]{gradshtein2000table}. For the special case of exponential correlation, where the correlation matrix is defined as ${\pmb{\Sigma}}_{i,j}=1$ for $i=j$ and ${\pmb{\Sigma}}_{i,j}=\rho^{\left| i-j \right|}$ for $i\neq j$, (\ref{eq:arbitrary_gamma_PDF1}) reduces to \cite[Eq. (2)]{peppas2011multivariate}.

\subsection{Multivariate $\Gamma \Gamma$ Distribution with Arbitrary Correlation}
The multivariate $\Gamma \Gamma$ distribution can be derived from the product of two multivariate Gamma RVs. Let us assume that the Gamma RVs in one multivariate Gamma distribution are correlated but they are independent from the Gamma RVs of the other multivariate Gamma distribution.\footnote{This assumption is reasonable, since in most practical wireless channels the effects of fast fading and shadowing are usually independent from each other.} Moreover, let $u$ and $\omega$ be two independent Gamma RVs with PDFs $f_u(u)$ and $f_\omega(\omega)$, respectively. The $\Gamma \Gamma$ RV is denoted by $z=u\omega$.

\begin{them}\label{them:PDF}
The joint PDF of the multivariate $\Gamma \Gamma$ distribution with arbitrary correlation is given by
\begin{align}\label{eq:arbitrary_GG_PDF}
&{f_{\pmb{z}}}\left( {\pmb{z}} \right)= \frac{{{2^N}{{\left| \pmb{W} \right|}^m}}}{{{{\left[ {\Gamma \left( \beta  \right)} \right]}^N}\Gamma \left( m \right)}}\sum\limits_{{i_1}, \ldots ,{i_{N - 1}} = 0}^\infty  {{\left(\frac{\beta}{\Omega}\right)} ^{ \frac{{N\left( {m + \beta } \right)}}{2} + \sum\limits_{j = 1}^{N - 1} {{i_j}} }} \prod\limits_{j = 1}^N \frac{{z_j^{{\mu _j}} }}{{p_{j,j}^{ {\eta _j}}}}\notag \\
&\times{K_{2{\eta _j}}}\left( {2\sqrt {\frac{{\beta {p_{j,j}}}}{\Omega }{z_j}} } \right)
  \prod\limits_{n = 1}^{N - 1} {\left[ {\frac{{{{\left| {{p_{n,n + 1}}} \right|}^{2{i_n}}}}}{{{i_n}!\Gamma \left( {m + {i_n}} \right)}}} \right]},
\end{align}
where $\mu_j \triangleq \frac{m+\alpha_j+\beta}{2}-1$, $\eta_j \triangleq \frac{m+\alpha_j - \beta}{2}$ with ${ {\alpha_j}} = i_j$ for $j=1$, ${ {\alpha_j}} = i_{N-1}$ for $j=N$, and ${ {\alpha_j}} = i_{j-1} + i_j$ for $j=2,3,\ldots,N-1$. Also, $K_v(\cdot)$ denotes the modified Bessel function of the second kind with order $v$ \cite[Eq. (8.407.1)]{gradshtein2000table}.
\end{them}

\begin{IEEEproof}
The multivariate $\Gamma \Gamma$ distribution can be obtained as
\begin{align}\label{eq:multi_product}
{f_{\pmb{z}}}\left( \pmb{z} \right) = \int_0^\infty   \ldots  \int_0^\infty  {\prod\limits_{n = 1}^N {\omega _n^{ - 1}{f_{{u _n}}}\left( {\frac{{{z_n}}}{{{\omega _n}}}} \right){f_{\pmb{\omega}} }\left( \pmb{\omega}  \right) d \pmb{\omega} } },
\end{align}
where $\pmb{z} \triangleq [z_1, z_2, \ldots, z_N]$, with $z_n=u_n \omega _n$. Moreover, $f_{u_n}(\cdot)$ is the PDF of $u_n$ following a Gamma distribution with parameter $\beta$ given in (2), while $\pmb{u} \triangleq [u_1, u_2, \ldots, u_N]$ is a vector containing $N$ independent and identically distributed (i.i.d.) Gamma-distributed RVs with the same parameter $\beta$. By substituting (\ref{eq:gamma_pdf}) and (\ref{eq:arbitrary_gamma_PDF1}) into (\ref{eq:multi_product}) and using \cite[Eq. (3.471.9)]{gradshtein2000table}, \eqref{eq:arbitrary_GG_PDF} can be derived.
\end{IEEEproof}

To the best of the authors' knowledge, \eqref{eq:arbitrary_GG_PDF} is new and has not been presented in previous literature. Although \eqref{eq:arbitrary_GG_PDF} is given in terms of infinite series, only a finite number of terms is needed to get a satisfactory accuracy (e.g., smaller than $10^{-6}$) for all considered cases (a detailed discussion is provided in Section \ref{se:5}).

\begin{coro}\label{lemm:CDF}
The joint CDF of the multivariate $\Gamma \Gamma$ distribution $\pmb{z}$ with arbitrary correlation is given by
\begin{align}\label{eq:arbitrary_GG_CDF}
&{F_{\pmb{z}}}\left( \pmb{z} \right) =   \frac{{{{\left| \pmb{W} \right|}^m}}}{{{{\left[ {\Gamma \left( \beta  \right)} \right]}^N}\Gamma \left( m \right)}}\sum\limits_{{i_1}, \ldots ,{i_{N  -  1}}  = 0}^\infty  \prod\limits_{j = 1}^N p_{j,j}^{\! - \!m \!-\! {\alpha _j}}\notag \\
&\times G_{1,3}^{2,1}\left[ {\left. {\frac{{\beta {p_{j,j}}}}{\Omega }{z_j}} \right|\begin{array}{*{20}{c}}
1\\
{m\! + \!{\alpha _j},\beta ,0}
\end{array}} \right]\prod\limits_{n = 1}^{N - 1} {\left[ {\frac{{{{\left| {{p_{n,n \!+\! 1}}} \right|}^{2{i_n}}}}}{{{i_n}!\Gamma \left( {m \!+\! {i_n}} \right)}}} \right]},
\end{align}
where $G[\cdot]$ is a Meijer's-$G$ function \cite[Eq. (9.301)]{gradshtein2000table}.
\end{coro}

\begin{IEEEproof}
{We start with the definition of the CDF as
\begin{align}
&{F_{\pmb{z}}}\left( {\pmb{z}}\right) = \frac{{{2^N}{{\left| \pmb{W} \right|}^m}}}{{{{\left[ {\Gamma \left( \beta  \right)} \right]}^N}\Gamma \left( m \right)}}\sum\limits_{{i_1}, \cdots ,{i_{N \!-\! 1}}  =  0}^\infty  {{\left( {\frac{\beta }{\Omega }} \right)}^{\frac{{N\left( {m \!+\! \beta } \right)}}{2} +  \sum\limits_{j = 1}^{N \!-\! 1} {{i_j}} }}\notag \\
&\times\prod\limits_{n = 1}^{N \!-\! 1} {\left[ {\frac{{{{\left| {{p_{j,j + 1}}} \right|}^{2{i_j}}}}}{{{i_n}!\Gamma \left( {m + {i_n}} \right)}}} \right]} \prod\limits_{j = 1}^N { {\int_0^{{z_j}} {\frac{{z_j^{{\mu _j}}}}{{p_{j,j}^{{\eta _j}}}}{K_{2{\eta _j}}}\left( {2\sqrt {\frac{{\beta {p_{j,j}}}}{\Omega }{z_j}} } \right)d{\pmb{z}}} }}. \notag
\end{align}
By transforming the Bessel function ${K_v}\left( \cdot \right)$ into a Meijer's-$G$ function through \cite[Eq. (14)]{adamchik1990algorithm} and using \cite[Eq. (26)]{adamchik1990algorithm}, we can conclude the proof with  \cite[Eq. (9.31.5)]{gradshtein2000table}.}
\end{IEEEproof}

Note that for $(m+\alpha_j-\beta) \notin \mathbb{Z}$ and by using \cite[Eq. (07.34.26.0004.01)]{Wolfram2013function}, \eqref{eq:arbitrary_GG_CDF} can be written in terms of the more familiar generalized hypergeometric function ${}_1{F_2}\left( \cdot \right)$  as
\begin{align}
&{F_{\pmb{z}}}\left( \pmb{z} \right)= \frac{{{{\left| {\pmb{W}} \right|}^m}}}{{{{\left[ {\Gamma \left( \beta  \right)} \right]}^N}\Gamma \left( m \right)}}\sum\limits_{{i_1}, \cdots ,{i_{N \!-\! 1}} = 0}^\infty  \prod\limits_{n  =  1}^{N \!-\! 1} {\left[ {\frac{{{{\left| {{p_{n,n \!+\! 1}}} \right|}^{2{i_n}}}}}{{{i_n}!\Gamma \left( {m \!+\! {i_n}} \right)}}} \right]} \notag \\
&\times\prod\limits_{j  = 1}^N p_{j,j}^{ \!-\! m \!-\! {\alpha _j}} \Bigg( \frac{{\Gamma \left( {\beta  \!-\! m \!-\! {\alpha _j}} \right)}}{{m \!+\! {\alpha _j}}}{{\left( {\frac{{\beta {p_{j,j}}}}{\Omega }{z_j}} \right)}^{m + {\alpha _j}}}\notag \\
&\times{}_1{F_2}\left( {m \!+\! {\alpha _j};1 \!-\! \beta  \!+\! m \!+\! {\alpha _j},1 \!+\! m \!+\! {\alpha _j};\frac{{\beta {p_{j,j}}}}{\Omega }{z_j}} \right)  \notag \\
&+ \frac{{\Gamma \left( {m \!+\! {\alpha _j} \!-\! \beta } \right)}}{\beta {{\left( {\frac{{\beta {p_{j,j}}}}{\Omega }{z_j}} \right)}^{-\beta} } } {}_1{F_2}\left( {\beta ;1 \!- \!m\! -\! {\alpha _j} \!+\! \beta ,1 \!+\! \beta ;\frac{{\beta {p_{j,j}}}}{\Omega }{z_j}} \right) \Bigg).\notag
\end{align}\vspace*{-2mm}

\begin{coro}\label{lemm:MGF}
The joint MGF of the $\Gamma \Gamma$ distribution $\pmb{z}$ with arbitrary correlation matrix is given by
\vspace*{-3mm}
\begin{align}\label{eq:arbitrary_GG_MGF}
&{M_{\pmb{z}}}\left( \pmb{s}\right) = \frac{{{{\left| \pmb{W} \right|}^m}}}{{{{\left[ {\Gamma \left( \beta  \right)} \right]}^N}\Gamma \left( m \right)}}\sum\limits_{{i_1}, \ldots ,{i_{N - 1}} = 0}^\infty  \prod\limits_{j = 1}^N p_{j,j}^{ - m - {\alpha _j}}  \notag \\
&\times G_{1,2}^{2,1}\left[ {\left. {\frac{{\beta {p_{j,j}}}}{{\Omega s_j}}} \right|\begin{array}{*{20}{c}}
1\\
{m + {\alpha _j},\beta }
\end{array}} \right]\prod\limits_{n = 1}^{N - 1} {\left[ {\frac{{{{\left| {{p_{n,n + 1}}} \right|}^{2{i_n}}}}}{{{i_n}!\Gamma \left( {m + {i_n}} \right)}}} \right]},
\end{align}
where $\pmb{s} \triangleq [s_1, s_2, \ldots, s_N]$.
\end{coro}

\begin{IEEEproof}
Based on the PDF expression (\ref{eq:arbitrary_GG_PDF}), the joint MGF of the multivariate $\Gamma \Gamma$ distribution is given by
\vspace*{-1mm}
\begin{align}\label{eq:GG_MGF}
{M_{\pmb{z}}}\left( \pmb{s} \right) =   \int_0^\infty   \ldots  \int_0^\infty  {\exp } \left( { - \sum\limits_{n = 1}^N {{s_n}{z_n}} } \right){{f_{\pmb{z}}}\left( \pmb{z} \right)d{\pmb{z}}}.
\end{align}
\vspace*{-2mm}

Following a similar procedure as in deriving (\ref{eq:arbitrary_GG_CDF}), we simply express the exponential function in terms of a Meijer's-$G$ function \cite[Eq. (11)]{adamchik1990algorithm} and utilize the relation \cite[Eq. (21)]{adamchik1990algorithm}. Then, the joint MGF expression in \eqref{eq:arbitrary_GG_MGF} is derived.
\end{IEEEproof}

\subsection{Special Cases}
Now we present some PDF and CDF expressions for special cases, e.g., bivariate arbitrarily correlated or exponentially correlated $\Gamma \Gamma$ RVs. Note that these expressions give the link to previous results and prove the correctness of our generalized formulas. For bivariate correlated $\Gamma \Gamma$ RVs, where ${\pmb{z}}=[z_1, z_2]$, the joint PDF and CDF reduce to \cite[Eq. (4)]{bithas2009bivariate} and \cite[Eq. (10)]{bithas2009bivariate}, respectively. For the special case of the exponentially correlated multivariate $\Gamma \Gamma$ distribution, (\ref{eq:arbitrary_GG_PDF}) simplifies to \cite[Eq. (5)]{peppas2011multivariate}. {Furthermore, when we consider independent $\Gamma \Gamma$ RVs, the joint PDF, CDF and MGF simplify to the product of multivariate independent $\Gamma\Gamma$ distributions \cite[Eqs. (2), (3) and (4)]{bithas2006performance}, respectively.} We finally point out that for exponentially correlated multivariate $\Gamma \Gamma$ RVs, the joint CDF formula (\ref{eq:arbitrary_GG_CDF}) reduces to \cite[Eq. (7)]{peppas2011multivariate} and the joint MGF expression (\ref{eq:arbitrary_GG_MGF}) reduces to \cite[Eq. (10)]{peppas2011multivariate}.

\section{Sum of $\Gamma \Gamma$ RVs with Arbitrary Correlation}\label{se:sum}

\subsection{PDF and CDF of the Sum of $\Gamma \Gamma$ RVs}
The sum of $N$ $\Gamma \Gamma$ RVs is defined as
\begin{align}\label{eq:sum}
S \triangleq \sum\limits_{n = 1}^N {{z_n}}  = \sum\limits_{n = 1}^N {{u_n}{\omega_n}},
\end{align}
where $u_n$ are i.i.d. Gamma RVs with parameters $(\beta, 1/\beta)$ whose PDF expression was given in \eqref{eq:gamma_pdf}, and $\omega_n$ are non-identical and correlated Gamma RVs with parameters $(m_n, \Omega_n)$ and joint PDF given in\cite[Eq. (11)]{karagiannidis2006closed}.
According to the approach presented in \cite{chatzidiamantis2011distribution}, we can rewrite \eqref{eq:sum} as
\begin{align}\label{eq:sum_ext}
S \!=\! \underbrace {\frac{1}{N}\sum\limits_{n  =  1}^N {{u_n}} \sum\limits_{n =  1}^N {{\omega _n}} }_{ \hat S} \!+\! \underbrace {\frac{1}{N}\sum\limits_{i  =  1}^{N \!-\! 1} {\sum\limits_{j = i  + 1}^N {\left( {{u_i} \!-\! {u_j}} \right)\left( {{\omega _i} \!-\! {\omega _j}} \right)} } }_{ \varepsilon},
\end{align}
where ${\hat S}$ is the approximation of the sum and $\varepsilon$ is the approximation error, respectively. The validity of this approximation has been investigated by Kolmogorov-Smirnov (KS) goodness-of-fit statistical test in \cite{chatzidiamantis2011distribution}. Then, ${\hat S}$ can be expressed as the product of two sums of Gamma RVs, i.e. $\hat S = {S_1}{S_2}$, where ${S_1}\triangleq \frac{1}{N}\sum\limits_{n = 1}^N {{u_n}} $ and ${S_2}\triangleq \sum\limits_{n = 1}^N {{\omega _n}}  $. It is well known that the sum of i.i.d. Gamma RVs $u_n$ remains Gamma distributed with parameters $(N\beta, 1/\beta)$ \cite[Eq. (7)]{al1985performance}. After variate transformation, we can conjecture that ${S_1}$ is Gamma distributed with parameters $(N\beta, 1/N\beta)$.

Following the similar method presented in \cite{karagiannidis2006closed}, let $\delta_n$ be the set of $N$ distinct eigenvalues of the $N \times N$ covariance matrix given by $\mathbf{K}_\mathbf{y}={\text{E}}(\mathbf{y}{\mathbf{y}^\dag })$, where
\begin{numcases}{{\text{E}}({y_{i,k}}{{y_{j,l}}})=}
\Omega_i/2, & if $i\!=\!j$ and $k\!=\!l$ \notag \\
\rho_{i,j}\sqrt{\Omega_i \Omega_j}/2, &if $i \!\neq \!j$ and $k\!=\!l $\notag \\
0, & otherwise. \notag
\end{numcases}

We assume that $\delta_n$ has algebraic multiplicity $\nu_n$, where $\nu_n/2 \in \mathbb{Z}$. Then, we can derive a closed-form PDF expression for ${S_2}$ as
\begin{align}
{f_{{S_2}}}\left( z \right) = \sum\limits_{i = 1}^N {\sum\limits_{j = 1}^{{m_i}} {{\Xi _N}\left( {i,j,\{ {m_n}\} _{n = 1}^N,\{ {\Omega _n}\} _{n = 1}^N} \right)} } {f_u}\left( {z;j,{\Omega _i}} \right),\notag
\end{align}
where ${f_u}\left( {z;j,{\Omega _i}} \right)$ is given in \eqref{eq:arbitrary_gamma_PDF1}, $m_n=\nu_n / 2$ should be integers, $\Omega _n=4\delta_n / \nu_n $, and
\begin{align}\label{eq:sum_gamma1}
&{\Xi _N}\left( {i,{m_i} \!-\! k,\{ {m_n}\} _{n \!=\! 1}^N,\{ {\Omega _n}\} _{n\! =\! 1}^N} \right)\!=\!
\frac{1}{k}\sum\limits_{\scriptstyle j,n\! =\! 1 \atop
\scriptstyle n \ne i }^N \frac{{{m_n}}}{{\Omega _n^j}} {{\left( {\frac{1}{{{\Omega _i}}}\! -\! \frac{1}{{{\Omega _n}}}} \right)}^{\! -\! j}}\notag \\
&\qquad \times{\Xi _N}\left( {i,{m_i} - k + j,\{ {m_n}\} _{n = 1}^N,\{ {\Omega _n}\} _{n = 1}^N} \right),
\end{align}
with $k=1,\ldots,m_i-1$ and
\begin{align}
&{\Xi _N}\left( {i,{m_i},\{ {m_n}\} _{n \!=\! 1}^N,\{ {\Omega _n}\} _{n \!=\! 1}^N} \right)\! =\! \frac{{\Omega _i^{{m_i}}}}{{\prod\nolimits_{t = 1}^N {\Omega _t^{{m_t}}} }}\prod\limits_{ j \ne i}^N {{{\left( {\frac{1}{{{\Omega _j}}} \!-\! \frac{1}{{{\Omega _i}}}} \right)}^{ \!-\! {m_j}}}}.\notag
\end{align}
With the PDF expressions of ${S_1}$ and ${S_2}$, we can derive the PDF of ${\hat S}$ as\footnote{Since the PDF expression involves the distinct eigenvalues of the covariance matrix $\mathbf{K}_\mathbf{y}$, it is not straightforward to derive similar expressions for the case of exponentially correlated $\Gamma\Gamma$ RVs. Yet, the authors in \cite{chatzidiamantis2011distribution} have given the PDF and CDF expressions for the case of independent $\Gamma\Gamma$ RVs.}
\begin{align}\label{eq:sum_gamma_pdf}
{f_{\hat S}}\left( z \right) &= \sum\limits_{i = 1}^N {\sum\limits_{j = 1}^{{m_i}} {{\Xi _N}\left( {i,j,\{ {m_n}\} _{n = 1}^N,\{ {\Omega _n}\} _{n = 1}^N} \right)} }  {f_{\gamma}}\left( {z;N\beta ,j,j{\Omega _i}} \right),
\end{align}
where $f_\gamma(\cdot)$ is the PDF of a $\Gamma \Gamma$ RV and is given by \cite[Eq. (2)]{bithas2006performance}
\begin{align}
{f_\gamma }\left( {z;N\beta ,j,j{\Omega _i}} \right) &= \frac{{2{{\left( {N\beta } \right)}^{\frac{{N\beta  + j}}{2}}}{z^{\frac{{N\beta  + j}}{2} - 1}}}}{{\Gamma \left( j \right)\Gamma \left( {N\beta } \right){{\left( {{\Omega _i} } \right)}^{\frac{{N\beta  + j}}{2}}}}} {K_{N\beta  - j}}\left( {2\sqrt {\frac{{N\beta z}}{{{\Omega _i} }}} } \right).
\end{align}
Finally, the corresponding CDF is given by
\begin{align}\label{eq:sum_gammagamma_cdf}
{F_{\hat S}}\left( z \right) &= \sum\limits_{i = 1}^N {\sum\limits_{j = 1}^{{m_i}} {{\Xi _N}\left( {i,j,\{ {m_n}\} _{n = 1}^N,\{ {\Omega _n}\} _{n = 1}^N} \right)} } \frac{1}{{\Gamma \left( j \right)\Gamma \left( {N\beta } \right)}}\notag \\
&\times G_{1,3}^{2,1}\left[\! {\left. {\frac{{N\beta }}{{{\Omega _i}}}z} \right|\begin{array}{*{20}{c}}
1\\
{N\beta ,j,0}
\end{array}}\!\right].
\end{align}
Note that the approximating PDF and CDF expressions \eqref{eq:sum_gamma_pdf} and \eqref{eq:sum_gammagamma_cdf} are both in analytical form, since they involve finite nested summations. The weights coefficients can be easily and quickly evaluated by using the recursive formula \eqref{eq:sum_gamma1}.

\subsection{Approximation Accuracy}
\begin{figure}[htbp]
\centering
\includegraphics[scale=0.6]{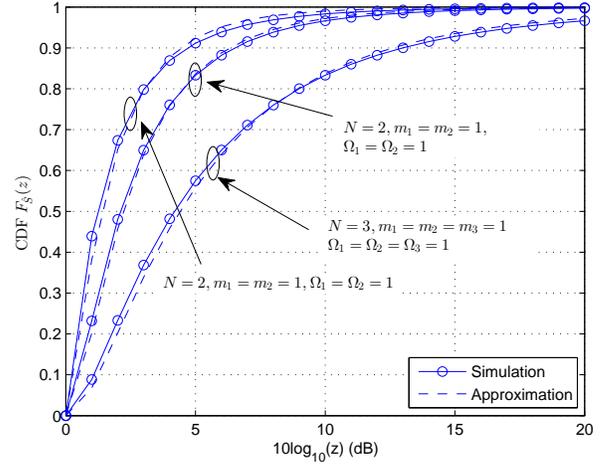}
\caption{Analytical and simulated CDF of the sum of $\Gamma \Gamma$ RVs ($\rho= 0.1$ and $\beta = 1$).
\label{fig:CDF_sum_gamma_gamma}}
\end{figure}

 {In order to evaluate the accuracy of our approximating result, the simulated and analytical \eqref{eq:sum_gammagamma_cdf} CDF curves of the sum of exponentially correlated $\Gamma \Gamma$ RVs are plotted in Fig. \ref{fig:CDF_sum_gamma_gamma}. Note that the analytical CDF expression can be only obtained by calculating the eigenvalues and algebraic multiplicity for each specific covariance matrix. From Fig. \ref{fig:CDF_sum_gamma_gamma}, it is clear that the analytical results approximate the exact ones with good accuracy. Moreover, note that the accuracy of the analytical CDF expression depends on the values of parameters $\rho$, $N$, $\beta$, $m_n$, and $\Omega_n$. As a general comment, higher values of $N$ and smaller values of $\Omega_n$ yield better accuracy.}

 \setlength{\textfloatsep}{10pt plus 1.0pt minus 2.0pt}
\setlength{\floatsep}{10pt plus 1.0pt minus 2.0pt}

Furthermore, we use a KS goodness-of-fit statistical test to validate the accuracy of the approximation results. The KS test statistic $T$ is defined as the maximal difference between the simulated and approximate CDFs. {The critical value $T_{\max}$ is given as $T_\text{max} \approx \sqrt{-\frac{1}{2v}\ln{\frac{\alpha}{2}}}$, where $\alpha = 5\%$ is the significance level and $v=10^4$ is the number of random samples of RVs \cite{papoulis2002probability}.} Then, $T_{\max}=0.0136$. The hypothesis $\mathbf{H}_0$ is not always accepted with $95\%$ significance and the accuracy of the proposed approximation depends on the combinations of the parameters $\rho$, $m_n$, $\Omega_n$, $\beta$ and $N$. For example, the hypothesis $\mathbf{H}_0$ is accepted with $T=0.0118<T_{\max}$ when considering exponential correlation with the following typical values, $m_1=m_2=1$, $\Omega_1=\Omega_2=1$, $\beta=1$, $\rho = 0.5$, and $N=2$. However, when we set $\rho = 0.1$ and keep other parameters the same, the hypothesis $\mathbf{H}_0$ is rejected with $T=0.0198>T_{\max}$.

\section{Applications in Wireless Communications}\label{se:4}
\subsection{Diversity Receivers}
In this section, the performance of various classes of multichannel diversity receivers, operating over arbitrarily correlated generalized-$K$ fading channels, is analyzed in terms of OP and BER. For each receive antenna, the transmitted signal is passing through the fading channel and perturbed by complex additive white Gaussian noise (AWGN) with zero mean and variance $N_0$. The instantaneous signal-to-noise ratio (SNR) of the baseband received symbol in the $n$th diversity branch is given by $\lambda_n = {\left| h_n \right|}^2 E_s/N_0$, where $E_s$ is the energy of the transmitted complex symbol and $h_n$ is the complex gain of the $n$th generalized-$K$ fading channel. By assuming that each fading channel has identical parameters, the average SNR per branch is the same and can be expressed as $\bar{\lambda}_n = {\text{E}}\left({\left| h_n \right|^2}\right) E_s/N_0 = \bar{\lambda}$,
where ${\text{E}}\left({\left| h_n \right|^2}\right)=m\Omega$. Furthermore, since shadowing occurs in large geographical areas, we assume that the shadowing parameters on each antenna are same, i.e., $\beta_n = \beta$, and the AWGN is uncorrelated among the diversity channels \cite{bithas2007diversity}. With $\lambda_n = {\left| h_n \right|}^2 \bar{\lambda}/m \Omega$, the joint PDF and CDF of $\pmb{\lambda} \triangleq [\lambda_1, \lambda_2, \ldots, \lambda_N]$ can be obtained as ${f_{\pmb{\lambda}} }\left( \pmb{\lambda}  \right) = {{{f_{\pmb{z}}}\left(  {\frac{{m\Omega }}{{\bar \lambda }}{\lambda _1},\frac{{m\Omega }}{{\bar \lambda }}{\lambda _2} \ldots ,\frac{{m\Omega }}{{\bar \lambda }}{\lambda _N}} \right)}}/{{{{\left( {\bar \lambda /m\Omega } \right)}^N}}}$ and ${F_{\pmb{\lambda}} }\left( {\pmb{\lambda}}  \right) = {F_{\pmb{z}}}\left(  {m\Omega } \frac{{\lambda _1}}{{\bar \lambda }} ,{ {m\Omega }\frac{{\lambda _2}}{{\bar \lambda }}} ,  \ldots ,{ {m\Omega } \frac{{\lambda _N}}{{\bar \lambda }}}  \right)$, respectively.

\subsubsection{Selection Combining}
The output SNR of SC diversity receivers is the highest instantaneous SNR among the multiple branches, i.e., $\lambda_{\text{sc}} = \max \{\lambda_1, \lambda_2, \ldots, \lambda_N\}$. An outage event happens when the SNRs of all branches fall below a given threshold $\lambda_{\text{th}}$. From \eqref{eq:arbitrary_GG_CDF}, the OP is obtained as
\begin{align}\label{eq:outage_sc}
&P_{\text{out}}(\lambda_{\text{th}})=  \frac{{{{\left| \pmb{W} \right|}^m}}}{{{{\left[ {\Gamma \left( \beta  \right)} \right]}^N}\Gamma \left( m \right)}}\sum\limits_{{i_1}, \ldots ,{i_{N - 1}} = 0}^\infty   \prod\limits_{n \!=\! 1}^{N \!-\! 1}{\left[ {\frac{{{{\left| {{p_{n,n \!+\! 1}}} \right|}^{2{i_n}}}}}{{{i_n}!\Gamma \left( {m \!+\! {i_n}} \right)}}} \right]} \notag \\
&\times G_{1,3}^{2,1}\left[ \!{\left. {\frac{{\beta m {p_{j,j}}\lambda _{\text{th}}}}{{\bar \lambda  }}  } \right|\begin{array}{*{20}{c}}
1\\
{m \!+\! {\alpha _j},\beta ,0}
\end{array}}\! \right]\prod\limits_{j = 1}^N p_{j,j}^{ - m - {\alpha _j}}.
\end{align}

The OP for small values of ${\lambda _\text{th}}$ is typically interesting in communication systems. We now focus on the high-SNR regime. Based on \cite[Eq. (9.303)]{gradshtein2000table} and \cite[Eq. (07.22.02.0001.01)]{Wolfram2013function}, the high-SNR outage can be derived from (\ref{eq:outage_sc}) and for $(m+\alpha_j-\beta) \notin \mathbb{Z}$ as
\begin{align}\label{eq:outage_sc_high}
&{P_{\text{out}}^\infty}(\lambda_{\text{th}})= \frac{{{{\left| \pmb{W} \right|}^m}}}{{{{\left[ {\Gamma \left( \beta  \right)} \right]}^N}\Gamma \left( m \right)}}\sum\limits_{{i_1}, \ldots ,{i_{N - 1}} = 0}^\infty  \prod\limits_{j = 1}^N p_{j,j}^{ - m - {\alpha _j}}\frac{{\Gamma \left( {{t_j} - {\tau_j}} \right)}}{{{\tau_j}}}\notag\\
&\times{{\left( {\beta m{p_{j,j}} \frac{{{\lambda _{\text{th}}}}}{{\bar \lambda }} } \right)}^{{\tau_j}}}
\prod\limits_{n = 1}^{N \!-\! 1} {\left[ {\frac{{{{\left| {{p_{n,n \!+\! 1}}} \right|}^{2{i_n}}}}}{{{i_n}!\Gamma \left( {m \!+\! {i_n}} \right)}}} \right]} \!+\! o\left({\lambda _{\text{th}}}^{\sum\limits_{j=1}^{N} \tau_j} \right),
\end{align}
where $\tau_j \triangleq \min(m+\alpha_j, \beta)$ and $t_j \triangleq \max(m+\alpha_j, \beta)$. \eqref{eq:outage_sc_high} indicates that using more antennas at the receiver affects the high-SNR slope, since the high-SNR slope grows linearly with $N$.

\subsubsection{Maximal-Ratio Combining}
The MRC receiver can achieve optimal performance in terms of maximizing the SNR at the output of the combiner \cite{simon2005digital}. The instantaneous SNR per symbol of an $N$-branch MRC receiver is given by $\lambda_{\text{MRC}}= \sum\nolimits_{n = 1}^N {{\lambda _n}} $. Then, the approximating expression for the OP of MRC receivers is given by
\begin{align}\label{eq:outage_mrc}
{P_{\text{out}}}\left( {{\lambda _{\text{th}}}} \right)  &\approx \sum\limits_{i = 1}^N {\sum\limits_{j = 1}^{{m_i}} {{\Xi _N}\left( {i,j,\{ {m_n}\} _{n = 1}^N,\{ {\Omega _n}\} _{n = 1}^N} \right)} }\notag \\
 &\times{F_\gamma }\left( {\lambda _{\text{th}};N\beta ,j,j{\Omega _i}} \right),
\end{align}
where ${F_\gamma }(\cdot)$ is given in \eqref{eq:sum_gammagamma_cdf}. {For $(N\beta-j) \notin \mathbb{Z}$, the OP in the high-SNR regime is given by
\begin{align}\label{eq:outage_mrc_high}
{P_{\text{out}}^\infty}\left( {{\lambda _{\text{th}}}} \right) &=
\sum\limits_{i = 1}^N {\sum\limits_{j = 1}^{{m_i}} {{\Xi _N}\left( {i,j,\{ {m_n}\} _{n = 1}^N,\{ {\Omega _n}\} _{n = 1}^N} \right)} } \notag \\
 &\times \frac{{\Gamma \left( {{e_j} - {c_j}} \right)}}{{{c_j}}} {{\left( {N\beta  \frac{{{\lambda _{\text{th}}}}}{{\Omega_i}} } \right)}^{{c_j}}} + o\left({\lambda _{\text{th}}}^{{c_j}} \right),
\end{align}
where $c_j \triangleq \min(N\beta, j)$ and $e_j \triangleq \max(N\beta, j)$. It is clear that the diversity order depends on the values of the parameters $N$, $\beta$ and $m_i$. For the case of MRC, $m_i$ in \eqref{eq:outage_mrc_high} should be integer.
}

We now consider the average BER of the MRC receiver for noncoherent binary frequency-shift keying (NBFSK). Based on \cite[Eq. (9.254)]{simon2005digital}, the average BER expression is given as $P_\textrm{b}^{\text{MRC}} = 0.5 {M_{{\lambda _{\text{MRC}}}}}\left( g \right)$, where $g = 0.5$ for NBFSK. Futhermore,  {the MGF of the output SNR can be obtained from (\ref{eq:arbitrary_GG_MGF}) as ${M_{{\lambda _{\text{MRC}}}}}\left( \pmb{s} \right) = {M_\lambda }\left( s,s,\ldots,s \right)$.}

{Following a similar method of \cite{peppas2011multivariate}, the approximate BER of MRC receivers for binary phase-shift keying (BPSK) modulation is given by
\begin{align}\label{eq:BER_MRC_BPSK2}
P_\textrm{b}^{\text{MRC}}   \approx \frac{1}{{12}}{M_{{\lambda _{\text{MRC}}}}}\left( 1 \right) + \frac{1}{4}{M_{{\lambda _{\text{MRC}}}}}\left( {\frac{4}{3}} \right).
\end{align}}
The approximate BER \eqref{eq:BER_MRC_BPSK2} provides a very useful metric of communication systems. Furthermore, \eqref{eq:BER_MRC_BPSK2} can reduce to \cite[Eq. (15)]{peppas2011multivariate} for exponentially correlated multivariate $\Gamma \Gamma$ RVs.

For the BER of BPSK in the high-SNR regime, we invoke the seminal parametrization in terms of diversity order $G_d$ and coding gain $G_c$. By keeping only the dominant term of (\ref{eq:BER_MRC_BPSK2}), the BER of MRC receivers over arbitrarily correlated generalized-$K$ fading channels is given by
\begin{align}\label{eq:BER_MRC_BPSK2_high}
P_\textrm{b}^{\textrm{MRC},\infty} = {\left( {{G_\textrm{c}} \times \Omega } \right)^{ - {G_\textrm{d}}}} + o\left( {{\Omega ^{ - {G_\textrm{d}}}}} \right),
\end{align}
where {
\begin{align}\label{eq:BER_MRC_BPSK2_order}
{G_\textrm{d}} & \triangleq \sum\limits_{j=1}^{N} \tau_j \\
{G_\textrm{c}} &\triangleq \Bigg(\frac{{{{\left| \pmb{W} \right|}^m}}}{{{{4\left[ {\Gamma \left( \beta  \right)} \right]}^N}\Gamma \left( m \right)}}\sum\limits_{{i_1}, \ldots ,{i_{N\! -\! 1}} \!=\! 0}^\infty  \prod\limits_{j \!= \!1}^N p_{j,j}^{ \!-\! m\! -\! {\alpha _j}}\Gamma \left( {{t_j} \!-\! {\tau_j}} \right) \Gamma \left( {{\tau_j}} \right)\notag \\
 &\times{{\left( {\frac{{3\beta {p_{j,j}}}}{{ 4}}} \right)}^{{\tau_j}}}\prod\limits_{n = 1}^{N - 1} {\left[ {\frac{{{{\left| {{p_{n,n + 1}}} \right|}^{2{i_n}}}}}{{{i_n}!\Gamma \left( {m + {i_n}} \right)}}} \right]}  \Bigg)^{\frac{-1}{\tau_j}}.
\end{align}}Based on the aforementioned expressions, we can notice that the diversity order does not depend on the correlation between fading channels, but depends on the shape parameter of small-scale fading and scale parameter of Gamma shadowing. Note that, for large values of $\beta$, the diversity order $G_d$ is limited by the shape parameter of small-scale fading.

\subsection{Free-Space Optical Communications}\label{se:FSO}

We consider a diversity FSO system with $N$ apertures and one beam, where the underlying sub-channels between pairs of transmit-receive apertures are correlated. Furthermore, we assume that the optimal combining (OC)\footnote{In FSO communications, the term OC is used instead of MRC.} scheme is deployed. Let $P_n$ be the fading coefficient of the $n$-th sub-channel, $n=1, 2, \ldots, N$, that follows the $\Gamma \Gamma$ distribution with arbitrary correlation. The instantaneous electrical SNR at the $n$-th receive aperture can be defined as $\lambda_n = (\eta P_n)^2/{N_0}$ and the average electrical SNR is $\bar{\lambda}_n = {|\eta {\text E}(P_n)|}^2/{N_0}$, where $\eta$ is the optical-to-electrical conversion coefficient. Without loss of generality, we normalize the average irradiance as ${\text E}(P_n)=1$ and assume that the average electrical SNRs at different receive apertures are equal, i.e., $\bar{\lambda}_n = \bar{\lambda}$. The variance of the noise in each receiver is ${N_0}/{2N}$ and the average BER is defined as \cite[Eq. (17)]{tsiftsis2009optical}
\begin{align}\label{eq:BER_OC}
P_\textrm{b}^{{\text{OC}}} = \int_0^\infty   \cdots  \int_0^\infty  Q \left( {\frac{\eta }{{\sqrt {2N{N_0}} }}\sqrt {\sum\limits_{n = 1}^N {P_n^2} } } \right){f_{\pmb{p}}}\left( {\pmb{p}} \right)d{\pmb{p}},
\end{align}
where $Q(\cdot)$ denotes the Gaussian $Q$-function \cite[Eq. (4.1)]{simon2005digital}, $\pmb{p} \triangleq [P_1, P_2, \ldots, P_N]$, and ${f_{\pmb{p}}}\left( {\pmb{p}} \right)$ is the joint PDF of the vector $\pmb{p} $ obtained with the aid of (\ref{eq:arbitrary_GG_PDF}).

In strong atmospheric channels, the fading parameters $\beta$ and $m$ can be respectively expressed as \cite[Eqs. (5)-(6)]{zhang2015unified}
\begin{align}
\beta  &= {\left( {\exp \left[ {\frac{{0.49\sigma _2^2}}{{{{\left( {1 + 0.18{d^2} + 0.56\sigma _2^{12/5}} \right)}^{7/6}}}}} \right] - 1} \right)^{ - 1}},\label{eq:beta_OC} \\
m &= {\left( {\exp \left[ {\frac{{0.51\sigma _2^2{{\left( {1 + 0.69\sigma _2^{12/5}} \right)}^{ - 5/6}}}}{{{{\left( {1 + 0.9{d^2} + 0.62{d^2}\sigma _2^{12/5}} \right)}^{5/6}}}}} \right] - 1} \right)^{ - 1}},\label{eq:m_OC}
\end{align}
where $\sigma _2^2 = 0.492C_n^2{k^{7/6}}{L^{11/6}}$ and $d = \sqrt {k{D^2}/4L} $ with $L$ being the distance between transmitter and receiver, $k$ is the optical wavenumber, $D$ is the aperture
diameter and $C_n^2 = 1.7 \times {10^{ - 14}}$ is the refractive-index structure parameter.

The integral in \eqref{eq:BER_OC} is very difficult to be evaluated in closed-form, so we use a simple and accurate exponential approximation for $Q(\cdot)$ \cite[Eq. (14)]{peppas2011multivariate} to derive an approximate result. Then, \eqref{eq:BER_OC} can be written as
\begin{align}\label{eq:BER_OC_appro}
P_\textrm{b}^{\text{OC}} &\approx \int_0^\infty   \cdots  \int_0^\infty  \Bigg( \frac{1}{{12}}\exp \left( { - \frac{\eta^2}{4 N {N_0}} \sum\limits_{n = 1}^N {{P_n^2}} } \right)\notag \\
 & + \frac{1}{4 }\exp \left( { - \frac{\eta^2}{3 N {N_0}} \sum\limits_{n = 1}^N {{P_n^2}} } \right) \Bigg){f_{\pmb{p}}}\left( {\pmb{p}} \right)d{\pmb{p}}.
\end{align}
Substituting (\ref{eq:arbitrary_GG_PDF}) into \eqref{eq:BER_OC_appro} and using \cite[Eqs. (11), (14)]{adamchik1990algorithm} and \cite[Eq. (21)]{adamchik1990algorithm}, we derive the average BER as
\begin{align}\label{eq:BER_OC_appro2}
P_\textrm{b}^{\text{OC}} \approx \frac{1}{{12}}\Lambda \left( {m,\beta ,\sqrt {\frac{{\bar \lambda }}{{4N}}} } \right) + \frac{1}{{4}}\Lambda \left( {m,\beta ,\sqrt {\frac{{\bar \lambda }}{{3N}}} } \right),
\end{align}
where
\begin{align}
&\Lambda \left( {m,\beta ,x} \right) \triangleq \frac{{{2^{N\left( {m \!+\! \beta  \!-\! 2} \right)}}{{\left| {{\pmb W}} \right|}^m}}}{{{{\left[ {\pi \Gamma \left( \beta  \right)} \right]}^N}\Gamma \left( m \right)}}\sum\limits_{{i_1}, \ldots ,{i_{N\! -\! 1}} = 0}^\infty  \prod\limits_{n = 1}^{N \!-\! 1} {\left[ {\frac{{{{\left| {2{p_{n,n \!+\! 1}}} \right|}^{2{i_n}}}}}{{{i_n}!\Gamma \left( {m \!+ \!{i_n}} \right)}}} \right]}\notag \\
&\times \prod\limits_{j \!=\! 1}^N p_{j,j}^{ \!-\! m \!- \!{\alpha _j}}G_{4,1}^{1,4}\left[ {\left. {{{\left( {\frac{{4x\Omega }}{{\beta {p_{j,j}}}}} \right)}^2}} \right|\begin{array}{*{20}{c}}
{\frac{{1\! -\! \left( {m \!+\! {\alpha _j}} \right)}}{2},\frac{{2 \!-\! \left( {m \!+\! {\alpha _j}} \right)}}{2},\frac{{1 - \beta }}{2},\frac{{2 \!-\! \beta }}{2}}\\
0
\end{array}} \right] .\notag
\end{align}

Note that for the special case of exponentially correlated atmospheric turbulence, \eqref{eq:BER_OC_appro2} coincides with a previously known expression \cite[Eq. (19)]{peppas2011multivariate}.

\section{Numerical Results}\label{se:5}

\subsection{RF Communication Systems}

\setlength{\textfloatsep}{10pt plus 1.0pt minus 2.0pt}
\setlength{\floatsep}{10pt plus 1.0pt minus 2.0pt}

\begin{figure}[!t]
\centering
\includegraphics[scale=0.6]{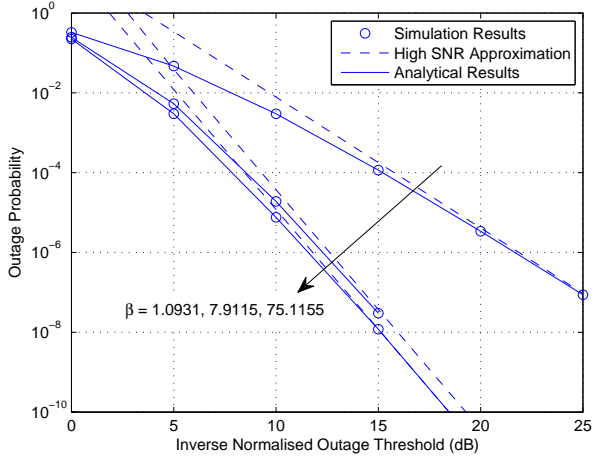}
\vspace*{-2mm}\caption{Analytical, simulated and high-SNR approximation OP of SC diversity receivers with three branches over exponentially correlated generalized-$K$ fading channels against the inverse normalized outage threshold ${{\bar \lambda  }}/{{\lambda _{\text{th}}}}$ ($\rho= 0.25$ and $m = 2$).
\label{fig:Outage_exp_high}}
\end{figure}

\begin{figure}[ht]
\centering
\includegraphics[scale=0.6]{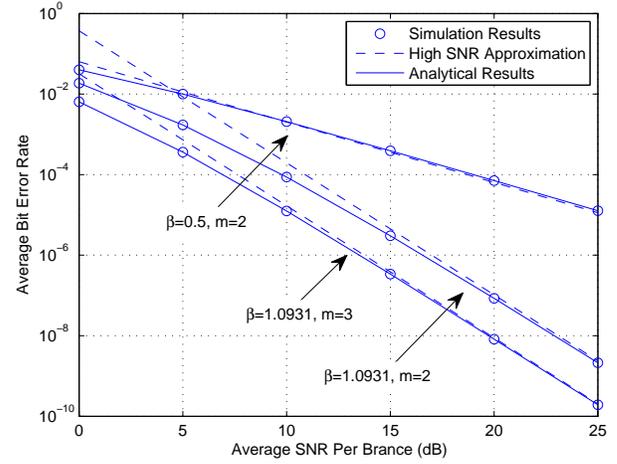}
\vspace*{-2mm}\caption{Average BER of triple SC diversity receivers over linearly correlated generalized-$K$ fading channels against the average SNR per branch $\bar{\lambda}$.
\label{fig5}}
\end{figure}

\begin{figure}[ht]
\centering
\includegraphics[scale=0.6]{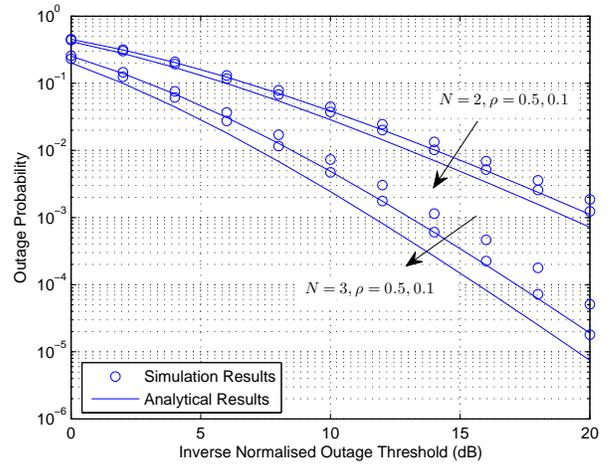}
\vspace*{-2mm}\caption{OP of MRC diversity receivers over exponentially correlated generalized-$K$ fading channels against the inverse normalized outage threshold ${{\bar \lambda  }}/{{\lambda _{\text{th}}}}$ ($\beta=1$, $m_n=m=1$).
\label{fig: Outage_MRC_exp_k111_m1_N2_N3}}
\end{figure}

\begin{figure}[ht]
\centering
\includegraphics[scale=0.6]{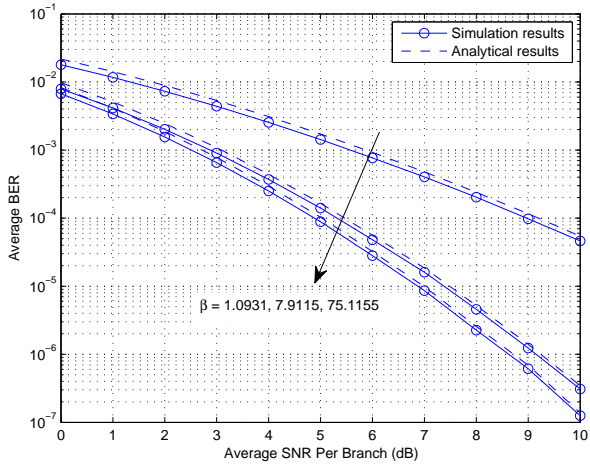}
\vspace*{-2mm}\caption{Bit error rate of quadruple branch MRC diversity receivers over exponentially correlated generalized-$K$ fading channels against the average SNR per branch $\bar{\lambda}$ ($m=2$, $\rho=0.3$).
\label{fig: BER_shadowing_exp_BPSK}}
\end{figure}

\begin{figure}[htp]
\centering
\includegraphics[scale=0.6]{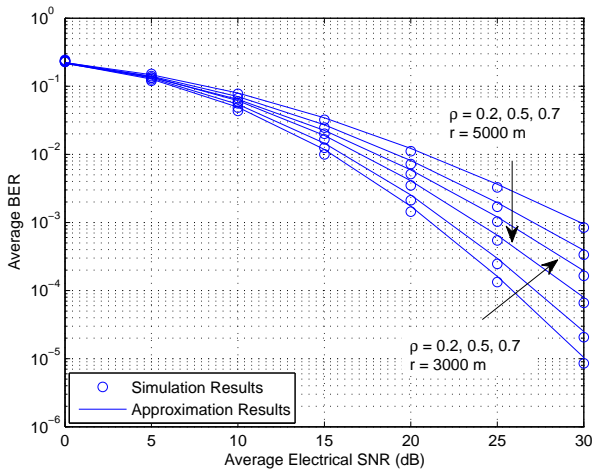}
\vspace*{-2mm}
\caption{Average BER of space diversity FSO system with three apertures and one
beam over strong turbulence fading channels against the average electrical SNR $\bar{\lambda}$.
\label{fig:FSO_BER_exp}}
\end{figure}

In our simulation, without loss of generality, it is assumed here that $I_1=I_2=\ldots=I_{N-1}$ and every branch has the same average SNR $\bar{\lambda}$. For the OP of three branches with SC receivers, the Monte-Carlo simulation results, analytical results (\ref{eq:outage_sc}) and high-SNR approximation \eqref{eq:outage_sc_high} are compared in Fig. \ref{fig:Outage_exp_high}. We again assume exponential correlation. The shadowing parameter $\beta$ corresponds to $1.0931$, $7.9115$ and $75.1155$ for frequent heavy shadowing, Karasawa shadowing and infrequent light shadowing, respectively \cite{peppas2011multivariate}. It is clear that for exponentially correlated generalized-$K$ fading channels, the analytical results agree closely with the exact OP results obtained via simulations. Furthermore, the high-SNR approximations are sufficiently tight and become exact, even at moderate SNR values. As expected, the OP that corresponds to light shadowing ($\beta = 75.1155$) is larger than the OP that corresponds to heavy shadowing ($\beta = 1.0931$). Moreover, this trend is more pronounced for smaller values of $\beta$.

Figure \ref{fig5} presents the simulated, analytical (\ref{eq:BER_MRC_BPSK2}) and high-SNR asymptotic results (\ref{eq:BER_MRC_BPSK2_high}) for the BER of triple SC diversity receivers over linearly correlated generalized-$K$ fading channels. The linearly correlated model is a more general case than the exponential one with a correlation Toeplitz structure matrix defined as ${\pmb{\Sigma}}_{i,j}=1$ for $i=j$ and ${\pmb{\Sigma}}_{i,j}={\pmb{\Sigma}}_{j,i}=\rho_{i,j}$ for $i\neq j$. We note that as the scale and shadowing coefficients increase, the average BER decreases. For all scenarios, the analytical results agree with the Monte-Carlo ones. Likewise, the diversity order and coding gains are accurately predicted. It is clear that the diversity order depends on the shadowing parameters.


The simulated and approximate OP of MRC receivers over exponentially correlated generalized-$K$ fading channels are investigated in Fig. \ref{fig: Outage_MRC_exp_k111_m1_N2_N3}. It is obvious that the proposed approximation is a lower bound of the simulated results. However, the difference is not greater than 2 dB in all cases. Furthermore, the approximation results are more accurate with the increased correlation coefficient $\rho$ and the decreased value of $N$.

{We present the simulated and approximate BER performance of quadruple MRC receivers over exponentially correlated generalized-$K$ fading channels in Fig. \ref{fig: BER_shadowing_exp_BPSK}. It is clear that the BER decreases as $\beta$ increases. Approximate BER curves are close to the simulated results and can be used as an upper bound. Moreover, the gap between two curves becomes smaller as the SNR increasing.}

\vspace*{-2mm}
\subsection{FSO Systems}

In Fig. \ref{fig:FSO_BER_exp}, the simulated and approximate (\ref{eq:BER_OC_appro2}) error performance of space diversity FSO links with $N=3$ receive apertures employing OC over exponentially correlated atmospheric turbulence channels, is depicted. We consider the case of the optical wavenumber $k=0.405 \times 10^7$, link distances $L=3000, 5000$m and correlated coefficients $\rho = 0.2, 0.5, 0.7$, respectively. Upon using \eqref{eq:beta_OC} and \eqref{eq:m_OC}, the respective values for $\beta$ and $m$ can be determined. From Fig. \ref{fig:FSO_BER_exp}, it is clear that the approximate results of cases under consideration closely agree with the ones obtained via Monte-Carlo simulations. {Note that the average BER increases by increasing $\rho$ and $L$ or decreasing $\bar{\lambda}$.}

\vspace*{-1mm}
\begin{spacing}{1}
\bibliographystyle{IEEEtran}
\bibliography{IEEEabrv,Ref}
\end{spacing}
\end{document}